# Characteristics of room temperature bipolar photoconductance in 150 GHz probe transients obtained from normal and irradiated silicon illuminated by 532 nm laser

Biswadev Roy, B. Vlahovic and Marvin H. Wu

Department of Mathematics & Physics, North Carolina Central University, Durham, N.C., 27707, U.S.A.

**Abstract**

A negative kink in excess conductivity is observed in p-type non-degenerate (moderate dopant concentration) silicon wafers when excited by a very narrow pulse of 532 nm laser appearing just after the complete positive decay of dark conductivity voltage. Most of the Si samples are pristine, and 3 of them are irradiated with gamma, proton, and chlorine ion beams respectively. These transients were examined using a time-resolved millimeter-wave conductivity apparatus (TRmmWC) and the radiofrequency (RF) voltage response (after laser cut-off) consistently reveals a positive peak with nominal decay to zero followed by a negative kink. This negative photoconductivity (NPC) kink develops just after the complete decay of the positive photoconductivity (PPC) and lasts typically ~ 36 μs. We present some data on general characteristics obtained from a set of normal (pristine doped Si) wafers and the gamma- and ion beam irradiated silicon (comparing with the parent pristine sample responses) for establishing possible new links that might enable estimation of defect parameters introduced in silicon.

## 1. Introduction

The probe beam in the photoexcited volume is absorbed by collisional interactions mainly due to the oscillatory charge carrier motion. Minority carrier density and its spatial distribution depend on the bulk and interface properties of the sample [1]. At millimeter-wave frequencies, we may neglect the effects of the drift of charge carriers and consider only charge-carrier generation, diffusion, and recombination. Carrier generation is a result of the pump excitation (laser) and the material whereas the mobility (μ) and carrier recombination steady-state lifetimes (τ) manifest from the interaction of the charge carriers and the material. The lifetime τ has 2 contributions, one from the recombination and trapping from the bulk ($\tau_b$) and another from the surface recombination velocity [2] that creates the decay channels ($\tau_s$) and relating them as $1/\tau = 1/\tau_b + 1/\tau_s$. In this paper, we will be dealing with $\tau_b$ as the data pertains only to the differential absorption of transmitted probe beam at 140, and 150 GHz when the silicon sample is excited using a 532 nm laser pump source. Such probe frequencies limit the effect of scattering in media compared to THz/FIR or higher frequencies [3].

Interestingly, the NPC signal does occur in the TRmmWC differential absorption signal collected using a negative voltage Schottky detector. The voltage transients in these signals resemble the same orders as found and reported in Fig. 7 of [4]. In n-type Czochralski Silicon NPC is reported by Yan Zhu et al. [5] and explained using the 2-defects model configured for 5 stages viz. equilibrium, under illumination, recombination, NPC, and recovery to equilibrium using an eddy-current-based probe for an n-type c-Si that is contaminated with Cr (no doping) for temperatures between 75-150$^0$ C and using Xenon flash lamp for excitation. In equilibrium, electrons have higher occupancy in *recombination-active defect* (RAD) than in *trapped-states* (TS) while under illumination. The positive peak in the transient under illumination is indicative of the increase in TS occupancy and when the pump source is cut off, the electron capture rate in RAD states is much higher than its emission rate from TS. This means for the p-type Si majority carrier (hole) concentration remains higher than the electron. This is followed by a slow TS emission of electrons back to the conduction band (CB) when the voltage is seen to be zero again.

NPC is reported by many authors as mentioned in ref. 5. Gold-doped Ge, Si, Co-doped Si, diamond, InN, $MoS_2$, GaAs quantum wells, and $Cd_xFe1-xSe$ system have exhibited NPC [5]. NPC is also observed in very thin transition dichalcogenides ($MoS_2$) [6]. They have reasoned it as a trion-induced effect (interaction of photo-induced electron-hole pairs with doping-induced charges) leading to a drastic reduction of conductivity. Barick et al. [7] have argued that when the sample is illuminated there occurs a rearrangement of electrons between the accumulation layer and the donor-like surface states which in turn leads to a reduction of electron concentration in the accumulation layer. This phenomenon can lead to less absorption of the passing microwave energy as a result of which, the detector will register negative voltage (more microwave power incident on it). Tavares et al. [8] reported NPC effect was observed in a p-type $Pb_{1-x}Sn_xTe$ film using blue LED light as a source of illumination for temperatures varying between 300K and 85K and they related the NPC effect directly to the defect states located in bandgap which act as trapping levels. NPC in heavily doped Si nanowire (NW) field-effect transistors (FET) was studied by Baek et al. [9]. They point out that the reduction in channel conductivity leading to NPC signal in Si NW is caused by the hot electron trapping by dopant ions and interfacial states. They have also argued that the changes observed in channel current are due to carrier trapping that causes NPC and photogeneration that results in PPC. Room temperature NPC also occurs in group III-V semiconductors such as in 0.6-0.7 eV bandgap n-type InN (indium-nitride) as reported by Wei et al. [10]. They used 532 nm laser stimulus and temperature-dependence results are published from using data from Hall measurement. They adopt an energy band model to explain PPC and NPC occurrences at a temperature greater and less than 160K for InN (a degenerate model with oxygen impurity) with high carrier concentration and designated quasi-Fermi levels for electrons and holes, and a recombination center.

This is the first time we are reporting the observation of NPC for p-type c-Si observed at 140, and 150 GHz probe frequency using the TRmmWC responses. In our experimental analysis, we have dealt with identifying some of the characteristics of the simultaneously occurring positive (PPC) and negative photoconductivity (NPC), due to reverse channel conductivity occurring in the sample after complete recombination of excited electrons following the complete decay of the PPC signal. We generally work with the PPC signal for all practical purposes and all practical use. It has not been explored yet how to make use of the PPC signal in conjunction with the simultaneously occurring NPC signal for any practical use to infer the material properties at various probe frequencies. It is well known that the optoelectronic properties of semiconductors are altered when the semiconductor is exposed to ionizing nuclear radiation. We have noted using TRmmWC signal from gamma-, proton, and chlorine ion beam induced defected samples that point and cluster defects are introduced in the Si samples due to the non-ionizing energy loss mechanism. This leads to an increase in sample resistivity when the fluence of the radiating particle is optimum that leads to an increased number of defects, and in a tri-exponential fit of the PPC signal through the irradiated data, we note that the third time-constant is enhanced with an increase in radiation dose [11]. For these reasons, we have investigated the behavior of PPC and NPC in the TRmmWC transients obtained at 150 GHz for some of the moderate resistivity Si samples that are free from any radiation-induced defects and another set of the same wafers but irradiated by gamma, proton, and chlorine ion beams. Some response characteristics of the respective samples are inter-compared and discussed in this paper. Section 2 of the paper describes the experiment with some details about the samples we have used, a short description of the theory behind the TRmmWC experiment, and the experimental arrangement with the schematic of the apparatus. Section 3 describes the results inferred about the relationship between the PPC and NPC peak voltages and voltage-time flux $\int_0^T V(t)dt$ with sample resistivity/dopant concentration, laser fluence. Section 3 also describes the results of the comparison of the PPC and NPC decay time constants for the normal and irradiated silicon samples. We summarize and conclude the results from this experiment and data analysis in Section 4.

## 2. TRmmWC Experiment
### 2.1 Sample Description

Moderate resistivity samples are considered for this study. One n-type pristine (Si-2) and 8 p-type c-Si wafers are used for the analysis of the nature of TRmmWC signal obtained in the transmission mode of operation at 140 and 150 GHz respectively. Out of the 8 p-type wafers 5 of them are normally doped samples and 3 are irradiated with 1.0 MGy 1.2 MeV $^{60}$Co γ exposed for about 33 hours, 2 MeV H$^+$ beam with fluence $10^{12}$ cm$^{-2}$, and 0.75 MeV Cl$^{2+}$ beam with fluence $10^{12}$ cm$^{-2}$ respectively. Table I below provides the sample details of each sample. The table gives sample parameters and measured quantities for normal type (not irradiated) comprising the set Si-1, 2, 4, 7, 8, and 10 and for irradiated samples Si-11, 12, 13. The same Si wafer as in Si-1 was used for irradiation that gave rise to Si-11, 12, and 13, and, these irradiated samples including the pristine Si-1 were probed using 150 GHz from a backward wave oscillator source while the pump fluence was maintained using the neutral density filter size 1.5. Most of the normal samples Si-1, 4, 7, 8, and 10 were probed using IMPATT oscillator source operated at 140 GHz and transients were averaged while the laser fluence was delivered to the sample using neutral density (ND) filter size 1.6.

**Table I** gives the sample type, thickness (t), resistivity (ρ), transmission coefficient (T), the PPC and NPC peak voltages for each set of samples, and respective flux magnitudes from the voltage-time profile for PPC and NPC respectively. * denotes Neutral Density (ND) filter size used to control the laser fluence. Actual magnitudes of laser fluences in µJoules cm$^{-2}$ are given for the ND filter size in Table II. ** Si-1 signifies the normal parent wafer that is used to irradiate with ion beams to obtain Si-11, 12, and 13

| Sample | Type | Thickness (µm) | Resistivity (Ω-cm) from 4-probe | PVL dopant conc. (cm$^{-3}$) | $T_{140}$ GHz | $T_{150}$ GHz | $\Delta V_{PPC}$ at 1.5ND (V) | $\Delta V_{PPC}$ at 1.6ND* (V) | $\Delta V_{NPC}$ at 1.5ND (V) | $\Delta V_{NPC}$ at 1.6ND (V) | Flux PPC at 1.5ND (Wb) | Flux PPC at 1.6ND (Wb) | Flux NPC at 1.5 ND (Wb) | Flux NPC at 1.6 ND (Wb) |
|---|---|---|---|---|---|---|---|---|---|---|---|---|---|---|
| | | | | | | **Normal samples** | | | | | | | | |
| Si-1** | P | 530 | 15.13 | 8.923x10$^{14}$ | - | 0.64 | 0.0061 | - | -1.3x10$^{-4}$ | - | 1.3x10$^{-9}$ | - | -1.6x10$^{-9}$ | - |
| Si-2 | N | 525 | 50.05 | 8.88x10$^{13}$ | 0.66 | - | - | 0.0014 | - | -6.7x10$^{-5}$ | - | 5.5x10$^{-9}$ | - | -6.2x10$^{-10}$ |
| Si-4 | P | 525 | 16.88 | 7.987x10$^{14}$ | 0.64 | - | - | 0.0012 | - | -6.4x10$^{-5}$ | - | 5.3x10$^{-10}$ | - | -5.8x10$^{-10}$ |
| Si-7 | P | 500 | 23.69 | 5.672x10$^{14}$ | 0.67 | - | - | 7.5x10$^{-4}$ | - | -2.7x10$^{-5}$ | - | 1.7x10$^{-10}$ | - | -2.1x10$^{-10}$ |
| Si-8 | P | 725 | 127.56 | 1.045x10$^{14}$ | 0.61 | - | - | 5.3x10$^{-4}$ | - | -1.5x10$^{-5}$ | - | 9.6x10$^{-10}$ | - | -7.7x10$^{-11}$ |
| Si-10 | P | 530 | 40.74 | 3.285x10$^{14}$ | 0.77 | - | - | 0.0012 | - | -1.7x10$^{-4}$ | - | 1.6x10$^{-9}$ | - | -1.1x10$^{-9}$ |
| | | | | | | **Irradiated samples** | | | | | | | | |
| Si-11 (γ) | P | 530 | 15.84 | 8.518x10$^{14}$ | - | 0.60 | 0.0055 | - | -1.5x10$^{-4}$ | - | - | - | -8.6x10$^{-10}$ | - |
| Si-12 (H$^+$) | P | 530 | 150.01 | 8.88x10$^{13}$ | - | 0.54 | 0.0048 | - | -6.1x10$^{-5}$ | - | - | - | -2.5x10$^{-10}$ | - |
| Si-13 (Cl$^{2+}$) | P | 530 | 15.0 | 9.001x10$^{14}$ | - | 0.47 | 0.0012 | - | | - | - | - | -7.6x10$^{-11}$ | - |

**Table-II:** Provides the 86.4 mm$^2$ spot-size laser (pump) beam power (in milliwatts), fluence(µJ cm$^{-2}$), and the calculated photon density. The Intensity for all ND filter sizes can be converted to fluence using fluence at 0.0 ND, $I_0 = 10.1$ µJ cm$^{-2}$, $I = I_0/10^{ND}$

| ND Filter Size | 0 | 0.3 | 0.6 | 0.9 | 1.2 | 1.5 | 1.6 | 1.8 | 2.1 | 2.3 |
|---|---|---|---|---|---|---|---|---|---|---|
| Power (mW) | 22.14 | 12.31 | 5.93 | 3.29 | 1.63 | 0.91 | 0.63 | 0.44 | 0.14 | 0.09 |
| Fluence (µJ cm$^{-2}$) | 10.1 | 5.6 | 2.7 | 1.5 | 0.75 | 0.41 | 0.29 | 0.2 | 0.06 | 0.04 |
| Photon density (cm$^{-3}$) | 1.34x10$^{17}$ | 7.48x10$^{16}$ | 3.62x10$^{16}$ | 2.03x10$^{16}$ | 1.02x10$^{16}$ | 5.83x10$^{15}$ | 4.17x10$^{15}$ | 2.98x10$^{15}$ | 1.16x10$^{15}$ | 8.65x10$^{14}$ |

### 2.2 Apparatus Description

The free-space, quasi-optical apparatus TRmmWC system is well explained in Roy et al. [12]. The continuous-wave probe frequency was fixed at 150 GHz with a beam power of ~ 0.32mW. This apparatus uses pump-probe measurement employing a 532 nm laser as pump source and either a backward wave oscillator (BWO) operated in the frequency range 110-170 GHz or, a cavity-based IMPATT oscillator probe source operated at 140 GHz. The system detects photo conductance by differential absorption of microwave probe beam passing through the sample that is illuminated by the pulsed laser resulting in altered charge carrier density in the Si wafer within a penetration depth of ~ 1μm due to photoexcitation. This apparatus has 2 modes of measurement viz. transmission (T) mode and the reflection (R) mode respectively. The schematic of the instrument is given below in Figure 1. Ultrafast (345ps) pump laser fluences were varied between 0.53 to 10.6 μJ cm$^{-2}$ and RF transient data from TRmmWC were recorded for each instance. A highly sensitive Schottky zero-bias detector records the probe beam absorption as a function of delay after the pump source is cutoff. Transients were averaged 4096 times before storing the .csv files for each instance.

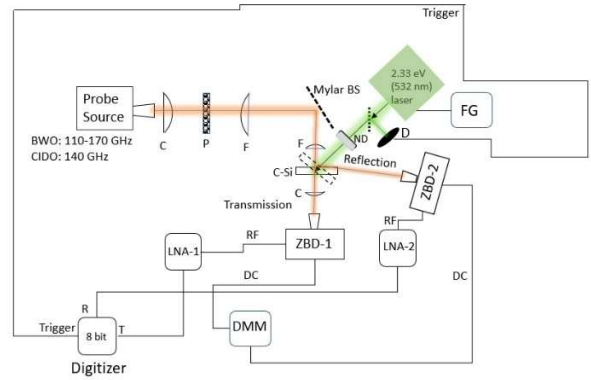

**Fig. 1:** Schematic of the operating TRmmWC apparatus with the arrangement of probe source (BWO/CIDO) and pump source (532 nm laser), function-generator for laser (FG), fast photodetector (D) for providing the trigger signal, and a variable neutral density filter (ND) with optical elements TPX collimator lens (C), Kapton grid polarizer (P), Mylar beam splitter (BS) reflecting 3.2% of the incoming beam from BWO/CIDO into the Si sample-path, TPX focus lens (F), the zero-bias Schottky diode detectors (ZBD-1 and 2), digital multimeter (DMM), low-noise amplifier (LNA) and the 8-bit digitizer. The RF and DC path is also shown. For these datasets, BWO frequency was fixed at 150 GHz and CIDO frequency is fixed at 140 GHz with beam power ~ 10 mW in both.

This results in RF voltage transients that can be tri-exponentially fitted to address 3 types of recombination processes [2].

The charge carriers in an optically stimulated sample give rise to complex conductance and the carriers tend to alter the dielectric properties of the material. The real part of the conductivity is related to the imaginary part of the dielectric constant of the semiconductor, and the imaginary part of conductivity is responsible to shift the resonance frequency of the probing microwave/millimeter-wave field [13]. Based on differential transmission principle (the same used for time-resolved microwave conductivity apparatus), TRmmWC response ( $\Delta V/V_0$ ) can be related to the charge conductance directly from using the sensitivity factor (K), charge mobility (μ) -carrier concentration ($n_c$), laser fluence (F) and the S/T ratio (S being multiple reflection parameter and T being the transmission parameter) products

$$\left(\frac{\Delta P}{P_T}\right)_{max} = -K\frac{\Delta V}{V_0} = -K\left[\frac{e(\mu_e+\mu_h)Z_0}{h\nu n}\left(\frac{S}{T}\right)\right]F \qquad (1)$$

ΔP being the change in probe beam power while laser illuminates the sample, $P_T$ is the power transmitted through the sample under dark (laser off) conditions. The power ratio can be related to the detector output voltage ratio using sensitivity K which was evaluated using a set of moderate resistivity silicon to be 0.2. The right-hand-side of Eq. (1 ) is the charge conductance involving the product of K and F with the ratio of the product of electronic charge e, the mobility of electron ($\mu_e$) and hole ($\mu_h$), $Z_0 =1/c\varepsilon_0 =376.73\Omega$ (the free-space impedance of air) to the product of Planck's constant (h), laser frequency (ν) and refractive index of silicon at 150 GHz (n=3.418). This ratio is multiplied by the (S/T) factor. Frequency-dependent reflection (ratio of sample reflected voltage to highly polished mirror voltages) and transmission coefficients (ratio of the DC voltage due to probe beam passing through the sample under dark conditions to the DC voltages registered by the same detector when probe beam is passing through free space) for the c-Si samples are obtained using the intensities of probe radiation incident on the sample and summed for the effect of multiple reflections at the entrance face (S), and for the net transmission through the plane-parallel etalon of thickness t (T) using equations 5(b) and 5(c) of reference [12]. For the Si-1, 4, and 11, DC reflection voltages were collected at 4 different probe frequencies (120, 130, 140, and 150 GHz) and the theoretical plots are shown along with the experimental reflection coefficients in Figure 2(a). Figure 2(b) shows only the theoretical transmission coefficients in the 110-170 GHz range for silicon. Transmission is highest at 170 GHz whereas, the surface reflection coefficient is maximum at 130 GHz. Keeping because of a modest amount of transmission and reflection, we have chosen 150 GHz as the probe frequency for analysis of the behavior of PPC and NPC signals for the normal and irradiated Si samples.

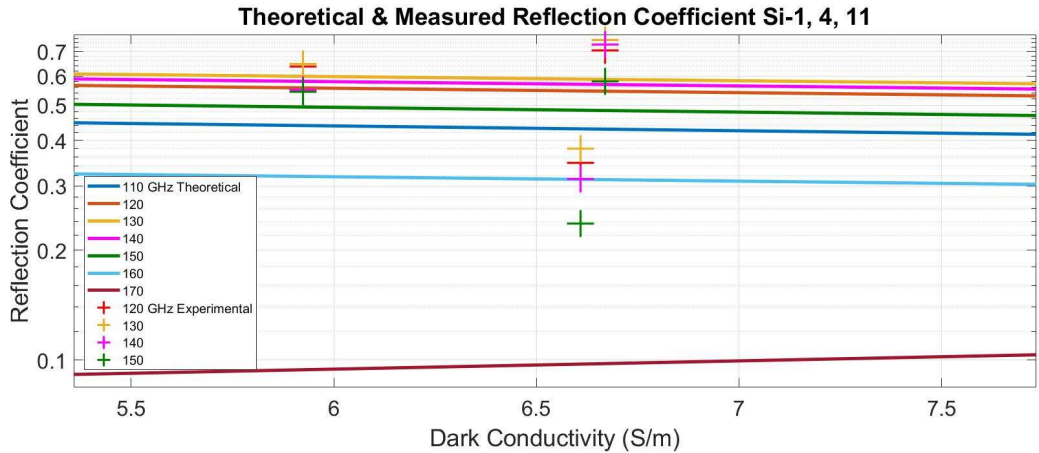

(a)

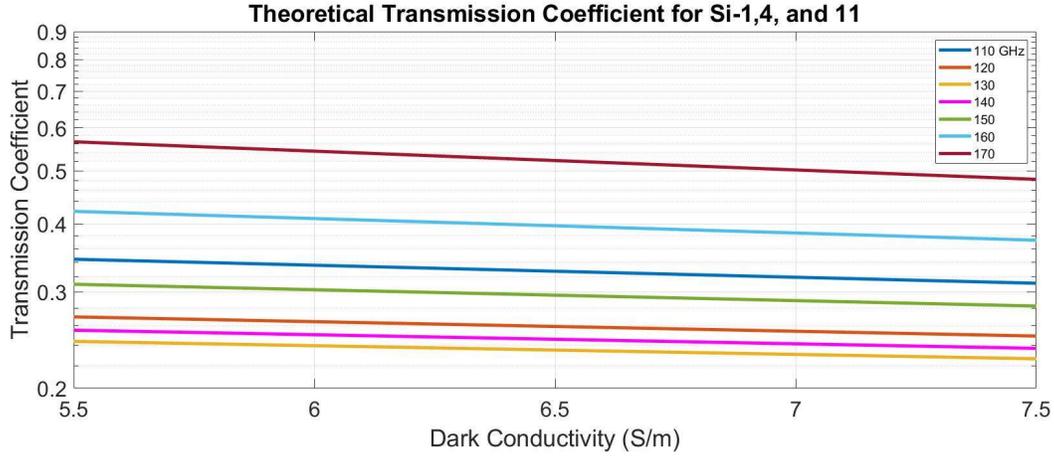

(b)

**Figure 2:** (a) shows the theoretically derived reflection coefficient as a function of dark conductivity (1/ρ) in the 110-170 GHz range (every 10 GHz) and the experimentally measured reflection coefficients when using the DC response of surface reflection of the samples and DC response of surface reflection from a highly polished mirror. (b) shows the theoretically calculated transmission coefficient through Si samples as a function of dark conductivity in the same range as 2(a) 5.5 to 7.5 Siemens/m.

## 3.0 Results and Discussion

3.1 PPC and NPC peaks in TRmmWC transients

Three separate wafers of the same parent (Si-1) with the same dopant concentration and resistivity were irradiated with gamma, proton, and chlorine beam respectively as mentioned in section 2.1. Our purpose is to compare the PPC and NPC characteristics obtained from irradiated samples with a complete set of normal Si samples with varying resistivity. Figures 3(a) and 3(b) shows the typical TRmmWC average transient response signal (in red) obtained from the normal c-Si sample at 150 GHz. The NPC (negative peak) for transmitted cases (Fig. 3c) and the reflected case (Fig. 3d) are shown as a function of the laser fluence for normal and irradiated Si samples respectively. These plots point out that the chlorine ion irradiated Si shows a greater NPC peak voltage for both transmission and reflection cases. Both the ion irradiated samples show a very gradual decrease of NPC peak with laser fluence compared to the pristine sample and the gamma-irradiated sample.

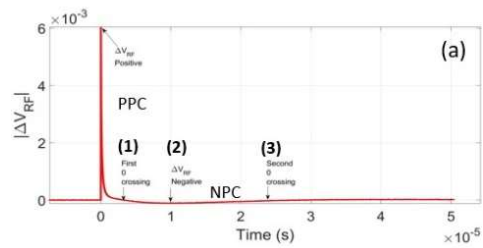
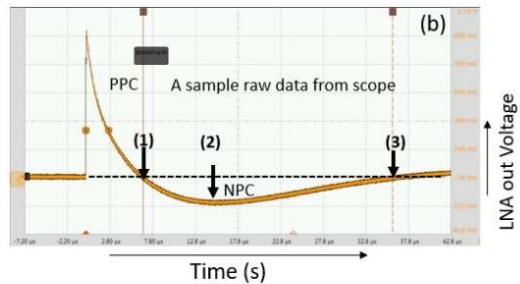
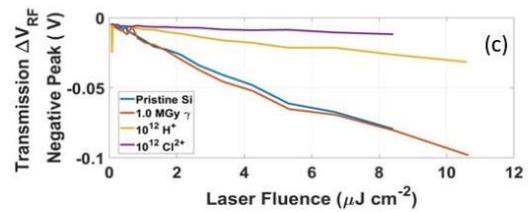
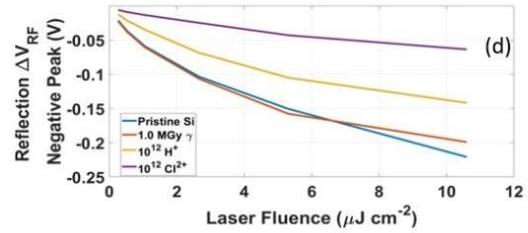

**Figure 3**: (a) Red curve is the typical 150 GHz RF transient voltage of one of these normal/irradiated silicon wafers as a function of delay. The positive photo-conductivity (PPC) peak occurs at around t=0 and the negative photo-conductivity (NPC) peak just after zero crossings at a delay of ~ 10 µs. (b) shows another transient (sample raw data) directly collected from the scope (single-shot) from either one of the Si samples to clarify the relevant PPC and NPC transients and points 1,2, and 3 mentioned in Fig. 3 (a), (c) Top right panel, shows the variation of this NPC (-ve peak) with laser fluence for RF acquisition in transmission (T) mode, and (d) the right bottom panel, is same as 3(c) but for NPC measured using the surface reflection (R) mode although, using the same set of irradiated Si wafers (both NPC peaks exhibit an increase in depth with laser fluence).

3.2 Variation of TRmmWC derived parameters with material properties, laser intensity

In this section, we have presented the 140 GHz and 150 GHz TRmmWC averaged transient derived parameters and compared those with the material properties such as resistivity, dopant concentration, dark conductance, and sample conductivity separately once for the normal samples and various laser intensity (fluence) bins, and again the similar analyses are carried out for the gamma, proton and chlorine ion beam irradiated silicon samples. Experimental data are acquired using the same free space geometry and the same set of quasi-optical arrangements. The TRmmWC signal derived parameters are the ratio, and differences of the PPC and NPC peak voltages, the fluxes (and their ratio) computed from the area under the PPC and NPC curves as they appear in the transients, the TRmmWC response ratio ($\Delta V/V_{dc}$), transmission coefficient (ratio of the transmitted microwave power (in voltage) to the incident probe beam power), the decay periods ($\tau$) important in the recombination dynamics of the normal and irradiated silicon samples.

*3.2.1 Variation of TRmmWC response-ratio with resistivity, flux, and dopant concentration*

To use a better quality of data we have used the laser fluences between 0.2 to 0.75 µJ cm$^{-2}$ (ND=1.3, 1.5, and 1.6). The ratio of the peak magnitudes of PPC and those for NPC are shown as a function of sample resistivity, flux, and PV Lighthouse (PVL) resistivity calculator derived dopant concentration (Figs. 4, 5, and 6 respectively). It is seen from Fig. 4 that the gamma- and chlorine ion does not introduce change in resistivity of the samples from Si-1 but, the resistivity increases up to about 10 times causing the peak PPC to NPC voltage ratio to jump from 50 (Si-1) to ~80. Figure 5 shows the 1.3 and 1.6 ND plots of PPC and NPC peak voltage TRmmWC response ratio. This figure consistently shows that the PPC peak voltage remains high at all dopant concentrations and the flux value as well follows the similar pattern as the PPC (high for higher fluence levels) except for the dopant concentration ~ 8x10$^{14}$ cm$^{-3}$ where we note that the PPC flux at lower fluence (1.6 ND) is higher than flux values obtained at 1.3ND. The insets shown in Fig. 5 consistently show the PPC and NPC peak voltage versus calculated flux values and saturation. There are 3 distinct strands also observed in the normal samples.

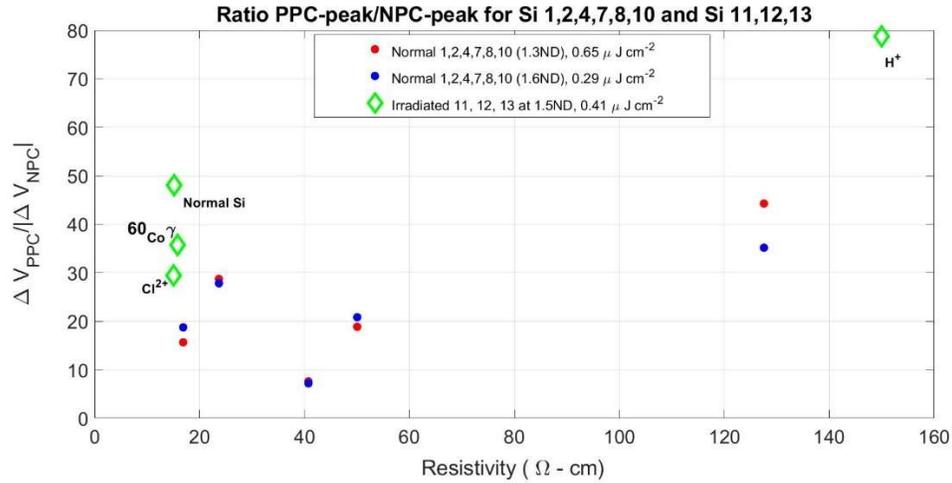

**Figure 4:** Shows the variation of the PPC peak voltage to NPC peak voltage ratio as a function of sample resistivity (4-probe). Notably, the normal samples yield the peak voltage ratio which is consistent with the laser fluence, however, the proton ion implanted sample introduces high resistivity and carrier traps that enhance the ratio greatly.

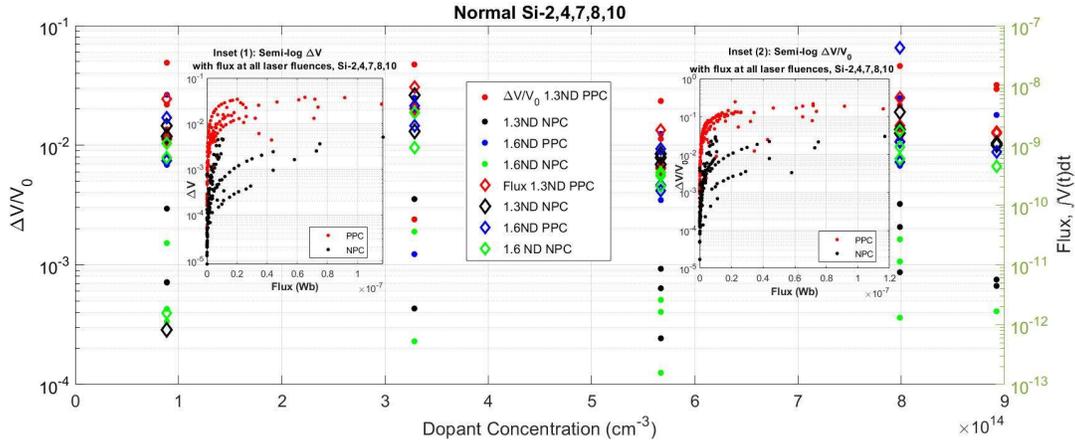

**Figure 5:** Primarily shows the variation of 1.3 and 1.6 ND laser fluence stimulated TRmmWC response ratio ($\Delta V/V_0$) and Flux with PVL calculator derived dopant concentration using 4-probe resistivity input 300K for normal Si samples 2,4,7,8 and 10. Left Inset: semi-log $\Delta V$ as a function of flux for PPC and NPC respectively, Right Inset: same as left inset except for the TRmmWC response ratio of the normal samples for PPC and NPC respectively.

Figure 6 shows the same TRmmWC signal-derived parameters and how they vary with sample parameters (PVL derived dopant concentration). It is evident from the inset semi-log plots that the peak response ($\Delta V$) and the TRmmWC response ratio (a photoconductivity measure) saturate with knees at about $10^{-8}$ Wb. The variation with flux also shows that for irradiated samples there are 2 different strands each for the PPC and NPC signals respectively. The response ratio for the irradiated samples also signifies that PPC signal response is about 1/100 of the NPC signal and it saturates faster than the NPC signal.

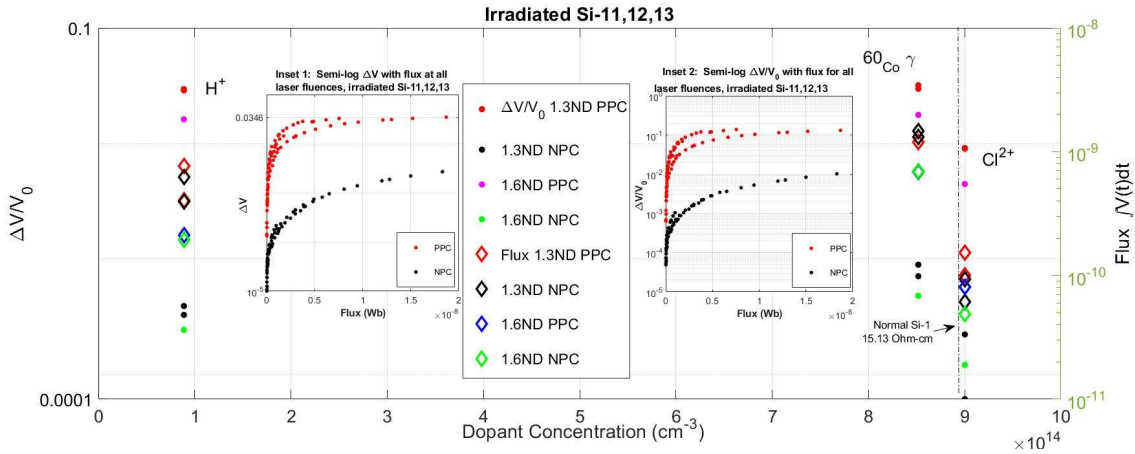

**Figure 6:** Same as in Figure 5, but, for the irradiated Si samples Si-11, 12, and 13

*3.2.2 Inter-dependence of normal and irradiated sample PPC and NPC peak voltages, and sensitivity of its ratio and differences to laser fluence and resistivity*

In transmission geometry of TRmmWC, the PPC voltage represents differential absorption (in time delays when the laser is suddenly switched off), and NPC occurring due to the existence of TS signifies a photo-resistive process in terms of its result leading to the passing of more RF signal through the sample (thereby increasing the negative voltage response). The reason we take the ratio of PPC and NPC peaks is to identify how the photo-conductance to its counterpart (photo-resistive) property changes with different laser stimulus magnitudes, and for a variety of sample resistivity and radiation-defected samples. The differences of the PPC and NPC peaks (magnitudes) provide the differences in carrier generation to trapping density in the sample for a given resistivity and laser fluence stimulus.

The existence of well-spaced ranges in laser fluence and sample resistivity provides an opportunity to explore the characteristics of TRmmWC derived PPC and NPC signals and how they vary with each other and also offers a scope to compare the responses between the normal samples and irradiated samples. Figure 7 shows the content of laser fluences that we used to acquire the TRmmWC data for all the samples taken together. The blue bars are for the fluences registered for normal samples Si-1,2,4,7,8,10 and the orange bars showing those for the irradiated samples Si-11,12,13.

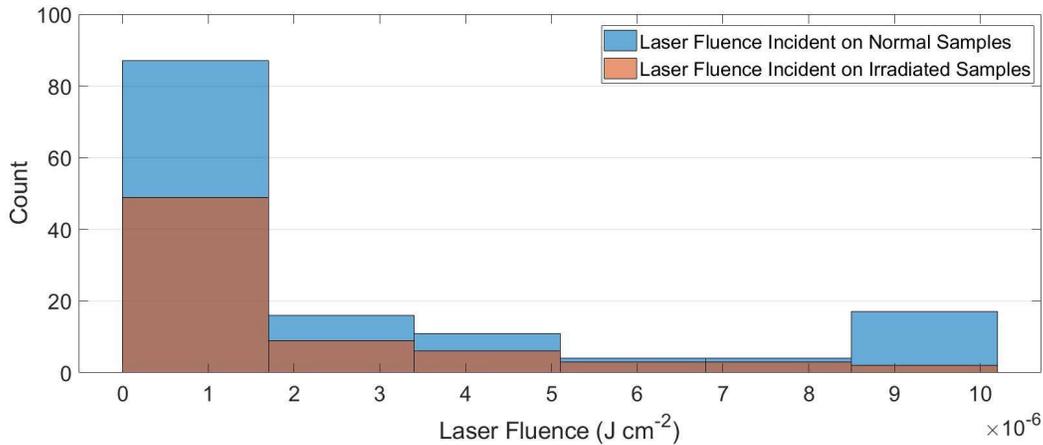

**Figure 7:** Shows the histogram of the laser fluence (counts) occurring in the experimental dataset used for this analysis.

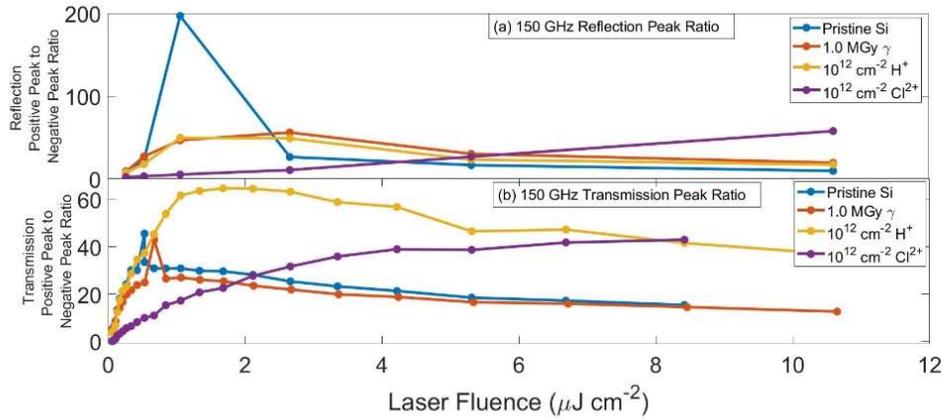

**Figure 8**:(a) top panel: shows the variation of positive to magnitudes of negative peak voltages for the reflected RF signals acquired at 150 GHz, these are shown for pristine sample, and one each for γ-, $H^+$ and $Cl^{2+}$ beam respectively. (b) bottom panel: same as in 8(a) except for the transmission mode from which RF signal is acquired. The correlation coefficient (for n=6) between the laser fluence and the ratio shown above are small -0.32, -0.16, -0.17 for pristine, γ- and $H^+$ irradiated Si respectively. It shows high for the $Cl^{2+}$ irradiated sample ~ 0.99.

In continuation of describing the NPC peak voltages and show how it relates to the laser fluence (shown in Figs. 3c and 3d), we show the peak ratios varying with laser fluences in Figures 8 (Fig. 8a showing the variation of reflection mode obtained PPC and NPC peak ratio and Fig. 8b showing the same for the transmission mode). The variation of the ratio of peak PPC (when electron occupancy in TS is very high) to the magnitude of NPC (dark-state) peak voltage (when hole concentration in TS is high due to rapid recombination of RAD with electrons from CB) as a function of laser fluences. It is evident that in reflection mode all the irradiated sample ratios vary similarly except for a bump in the normal sample (Si-1) peak ratio. However, proton irradiation causes the creation of traps inside Si that also give rise to an elevated TRmmWC NPC peak voltage, but, falls off slower compared to the normal (pristine/not irradiated) wafer response and the gamma exposed wafer responses respectively. Additionally, for the reflection case, we see a fair agreement of all 4 c-Si data except for pristine Si showing a peak at around 1 μJ cm.$^{-2}$ However, for the transmission case, we note that the proton damaged Si wafer exhibits a much higher peak ratio for laser fluence interval 1-4 μJ cm.$^{-2}$ It needs to be mentioned that transmission of the probe beam is through the bulk of the sample, but, with low penetration-depth of millimeter-wave (mmW) on Si causes the control volume to be very small compared to its thickness and does not show large differences in transmission coefficient, however, we see that the 2 MeV $H^+$ beam exhibit creation of large TS that trap carriers very effectively causing a spike in resistivity from 15 to 150 Ω-cm. An NPC model for this type of Cz-Silicon defects introduced by $Cl^{2+}$ radiation seems to be an important item since the radiation-damaged sample responds to mmW that highly correlates with laser fluence.

Plotting the NPC peak voltage as a function of the PPC peak voltages for the normal samples by showing data points using resistivity color codes (Fig. 9) and for the same, while showing the data points using color codes of the fluence intervals (Fig. 10), we note in general, there is an exponential dependence between the peak NPC and PPC voltages and they are identifiable by the resistivity of the irradiated (exposed) samples but, for the data points showing resistivity (Fig. 10) we see that except for the very low laser fluences, the data points are unevenly distributed in different NPC and PPC voltage ranges. Comparing Figs. 9 with 11 and Figs. 10 with 12, such a degree of unevenness in data is not shown in the irradiated sample plots (Fig. 11 for PPC versus NPC peak voltages shown by resistivity and Fig. 12 for the same shown by laser fluence).

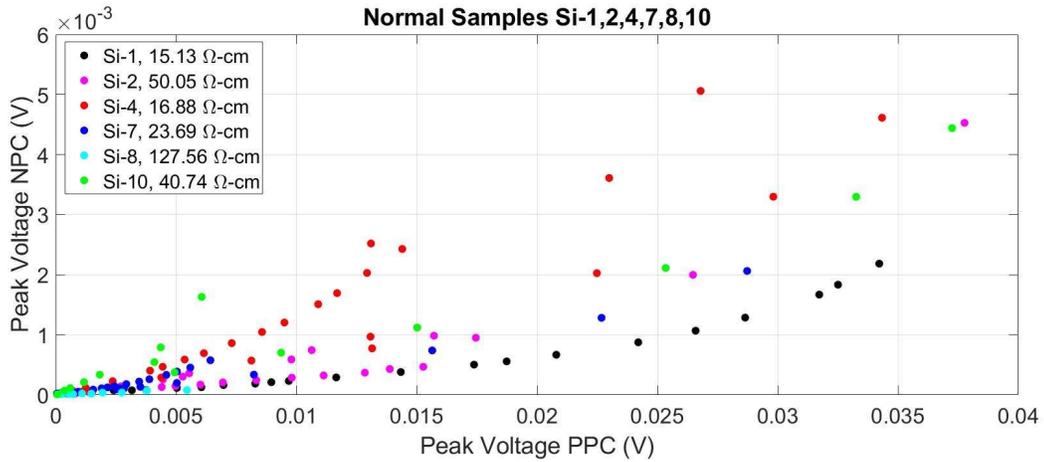

**Figure 9:** Plot of the NPC peak voltage (magnitude) as a function of PPC peak voltages for normal Si samples 1,2,4,7,8, and 10 data points shown by Si resistivity in Ohm-cm. Note the exponential dependence between the NPC and PPC peak voltages measured using probe frequency 150 GHz.

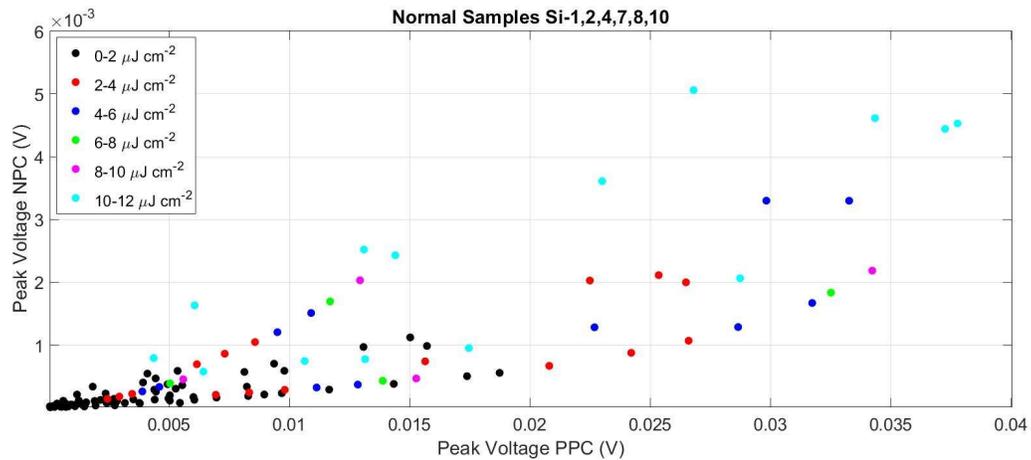

**Figure 10:** Same as in Fig. 9, but, for the plot showing data points labeled using the color-coding of the laser fluences used for stimulating the normal samples Si-1,2,4,7,8, and 10 respectively. We also note an exponential dependence between the NPC and PPC peak voltages obtained at probe frequency 150 GHz.

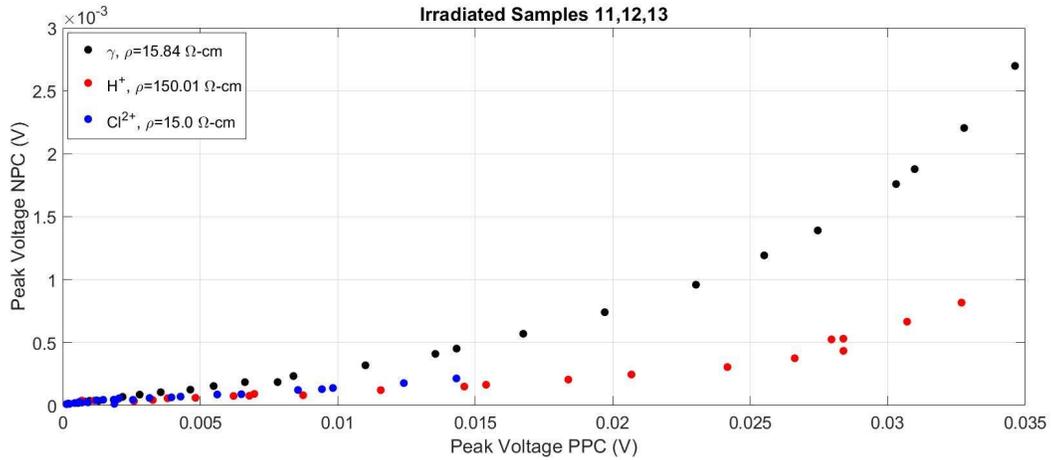

**Figure 11:** Same as shown in Figs. 9, and 10, except for the PPC and magnitudes of NPC peaks measured using irradiated Si-11,12, and 13 wafers at variable fluence by sample resistivity with probe frequency 150 GHz. Note: very distinct profiles are obtained for the 3 different resistivity samples.

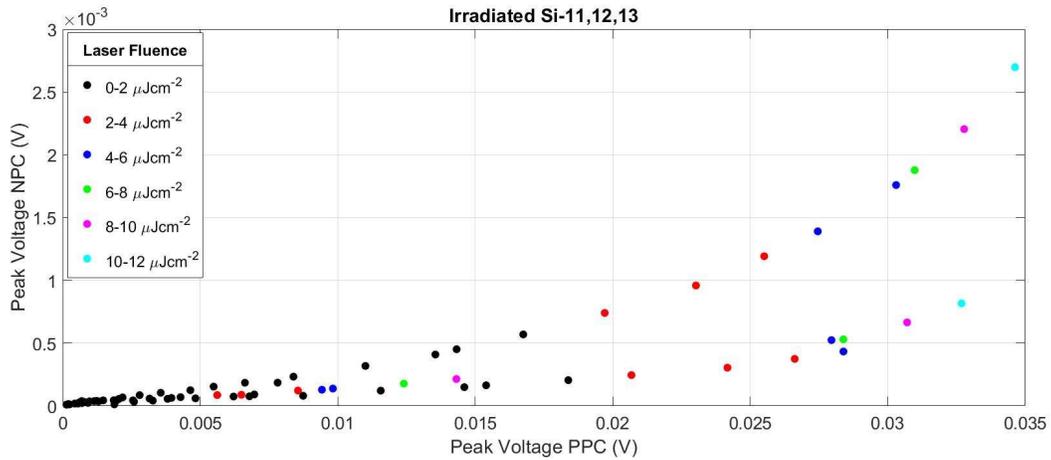

**Figure 12:** Same as in Fig. 11, except for the data points shown by laser fluence. Although the variation profile remains similar we note distinct laser fluence regimes in the dataset. There is an increase in the PPC and NPC peak with fluence as seen here.

Following the similar pattern of analysis as done before, we have looked at the variation of the peak voltages (PPC and NPC separately) for normal samples Si-1,2,4,7,8,10 as a function of laser fluence and labeling the data points by using the laser fluence range (ND filter sizes). These plots are shown in Figs. 13(a) and 13(b) respectively. Interestingly, it is found from Fig. 13(a) that the peak PPC voltages bear an exponential increase with laser fluence, whereas the peak NPC voltages bear almost a linear relationship with laser fluence.

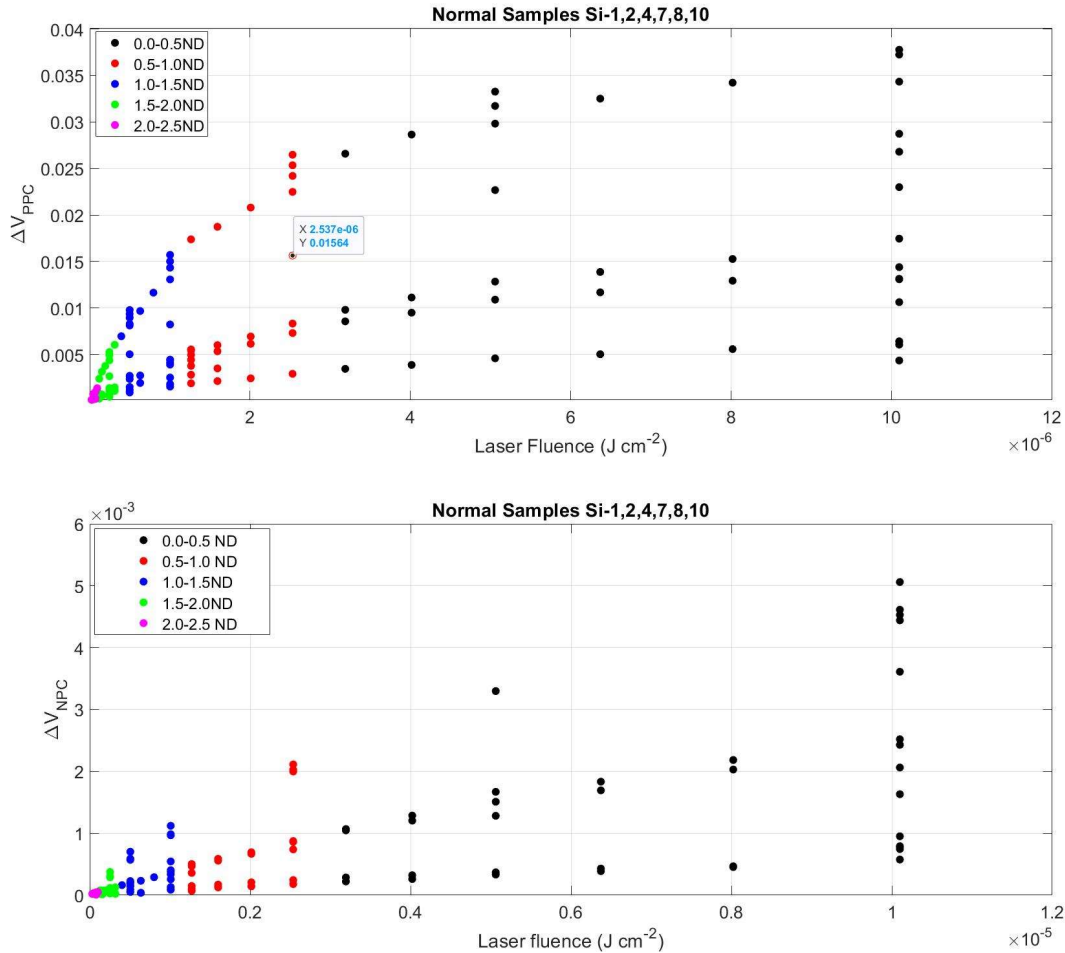

**Figure 13:** (a) shows the variation of the PPC peak voltages with laser fluence and (b) shows the magnitude of NPC peak voltage variation with laser fluence as well. The axis provides fluences in J cm$^{-2}$ however, the legend provides the ND filter size values associated with each color code used.

We have shown the differences between the PPC and NPC peak voltages observed in the irradiated samples from those registered by the parent wafer (Si-1) PPC and NPC peak voltages respectively in Figure 14. To reduce the complexity of this preliminary analysis, these plots are discussed for a fixed laser fluence (~0.41 µJ cm$^{-2}$, ND 1.5). It is seen that for the gamma-irradiated silicon sample the different voltages are clustered around 0 however, for the proton and chlorine ion beam irradiated Si there is strong interdependence with the proton beam sample showing a stronger dependence than the chlorine ion damaged sample. Plotting the ratio of the peak voltages with the laser ND filter size (reversed to represent laser fluence level) for the irradiated samples in Fig. 15 we note that the peak voltage ratio ($\Delta V_{PPC}/\Delta V_{NPC}$) becomes maximum at about 1.3ND filter size (~0.65 µJ cm$^{-2}$ fluence) and proton irradiated sample reaches maximum ratio ~150 at this fluence.

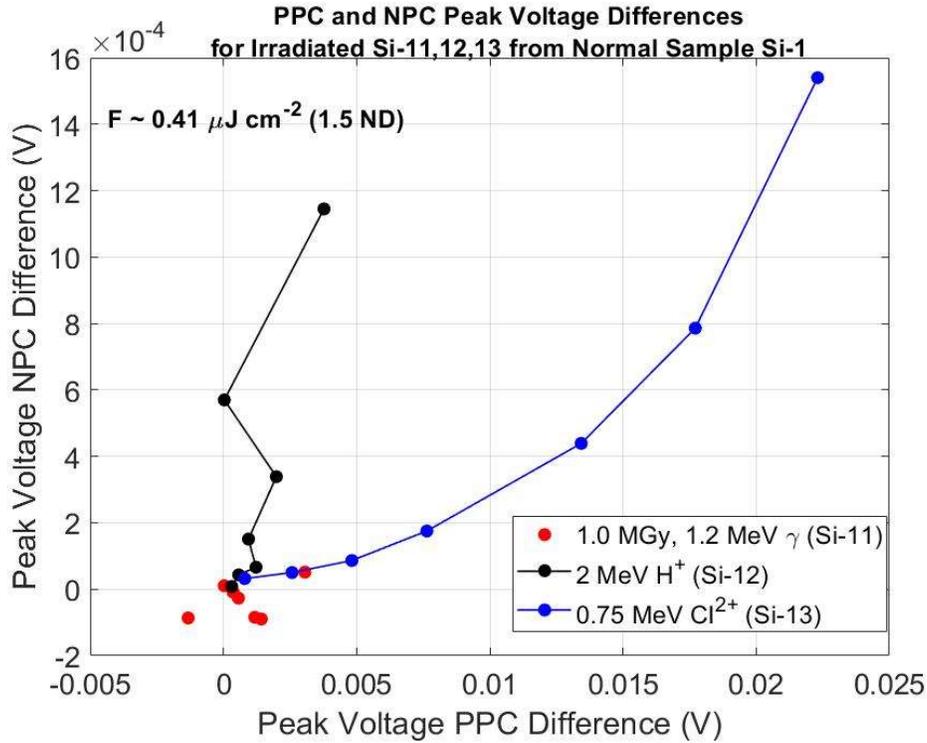

**Figure 14:** Departure of the gamma, proton, and chlorine ion irradiated silicon NPC and PPC peak voltages from the pristine (normal, Si-1) samples are plotted together for a fixed laser fluence ~0.41 µJ cm$^{-2}$. It is evident that for the gamma-irradiated sample, the peak voltage differences are clustered at a very low end. However, the 1.2 MeV proton and 0.75 MeV chlorine ion irradiated samples (Si-12, and 13) show a sharp rise in the profile.

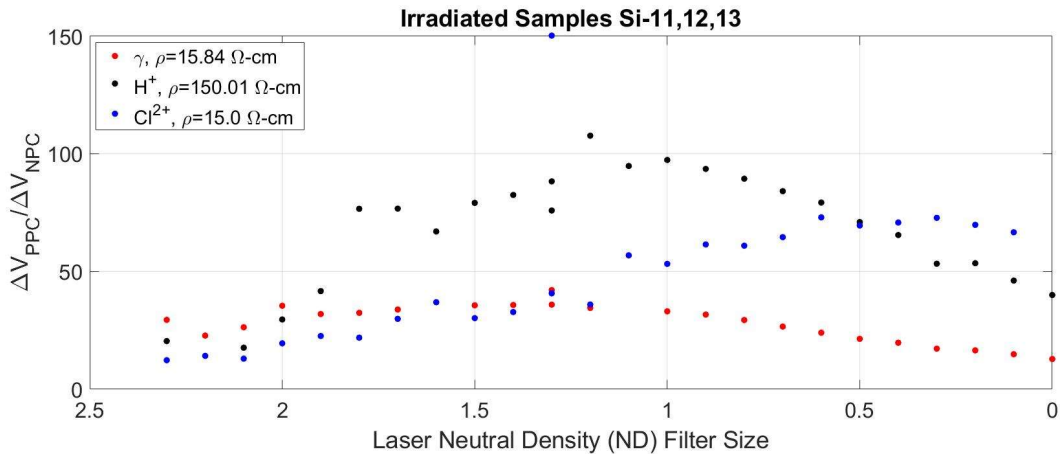

**Figure 15:** Irradiated sample ratio of maximum PPC to the magnitude of maximum NPC peak voltage; ratio peaks at ~1.3ND filter size but, the profile is very distinct for the 3 different irradiation types as seen here.

Further, we show the histogram of the laser neutral density filter (ND) sizes used during the experiment in Fig. 16(a) for the normal samples and also from the irradiated sample experiments. This is somewhat complementary to the histogram shown for laser fluences (Fig. 7). Using the normal samples data (Si-1,2,4,7,8, 10) we plot the ratio of the PPC and NPC peak voltages (Fig.

16b) and their differences (Fig. 16c) as a function of laser fluence both, by labeling the data points using color codes of sample resistivity. Once again, we note that for the high resistivity normal sample (Si-8) the peak ratio goes about in the 40-80 range and the differences between peak voltages for the normal samples exhibit higher PPC voltages for the low resistivity samples.

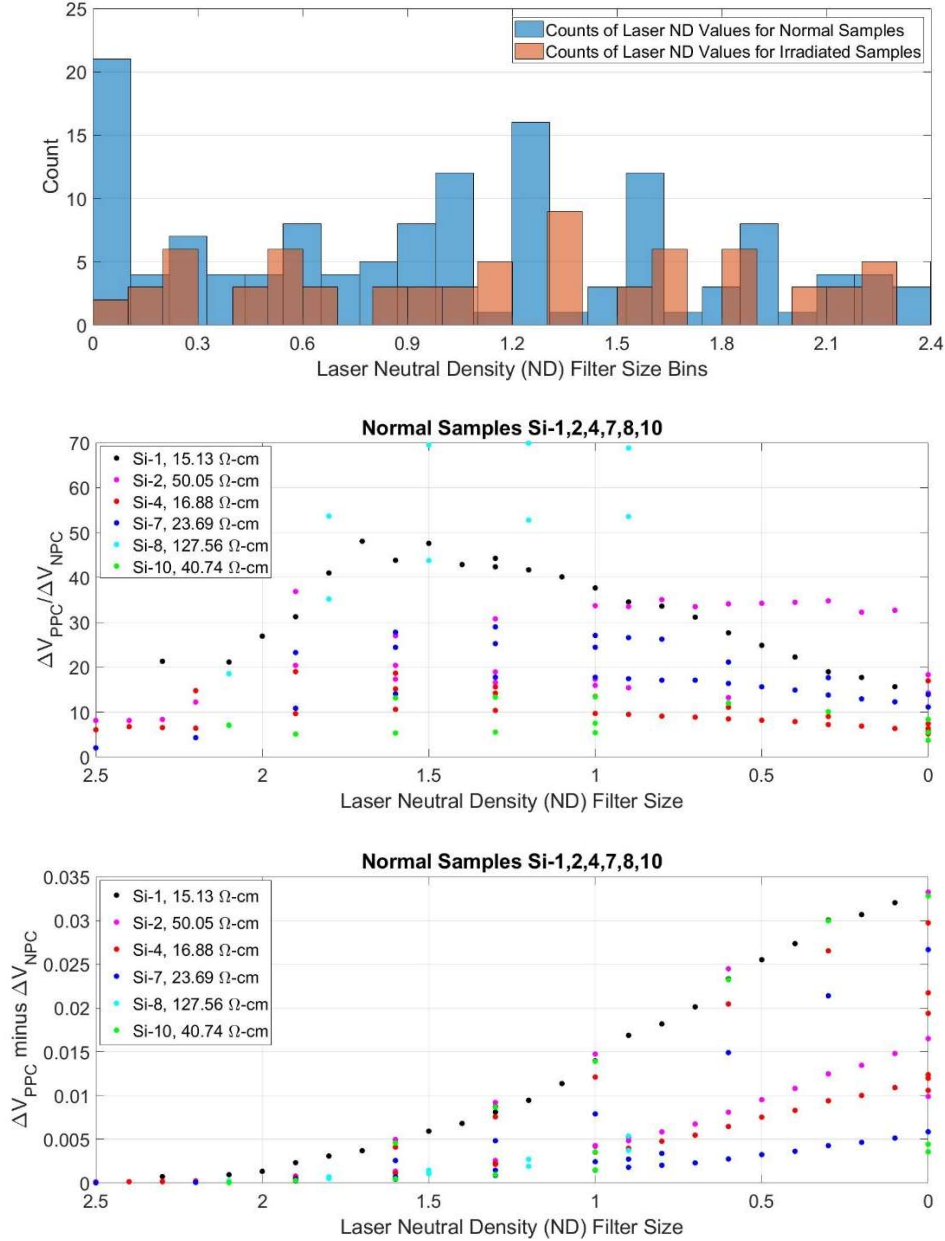

**Figure 16:** (a) Histogram of the laser neutral density (ND) filter sizes used during collection of the entire dataset for normal (blue bars) and irradiated samples (orange bars) respectively (refer to Table II for ND to intensity conversion). (b) shows the variation of the PPC to NPC (magnitude) peak voltage ratio with laser ND filter size, with color-coding of data points using resistivity values, and (c) shows the differences of the PPC and NPC (magnitude) peak voltages as a function of ND filter size with color-coding representing sample resistivity.

To visualize the joint variation of normal and irradiated sample peak PPC to peak NPC voltage ratio concerning laser intensity and samples resistivity we plotted the data in 3D using the laser intensity in the X-axis, the sample resistivity as the Y-axis, and the PPC to NPC peak voltage ratio as the Z-axis (Figure 17a) and again, plotted the differences between the peak PPC and NPC voltages for the irradiated and normal samples in one plot (Fig. 17b). It is characteristically seen that the peak voltage ratio for the irradiated samples peak very prominently at 1.3ND filter size($H^+$ irradiated Si), and very faintly at around 1.5ND filter size for the sample irradiated with $Cl^{2+}$ beam. The gamma irradiated sample shows a steady increase with laser fluence (or, with the decrease of laser ND filter size). Notably, $H^+$ irradiated sample shows a sharp rise in PPC/NPC peak voltage ratio and a sharp fall with an increase in laser intensity beyond 0.65 µJ cm.$^{-2}$ This behavior is in contrast to the differences in data variation shown in Fig. 17(b). For all samples, it is seen that the PPC and NPC voltage differences rise very gradually with laser fluence and $H^+$ irradiated sample shows much higher growth of PPC to NPC when compared with the normal sample (Si-8) with comparable resistivities.

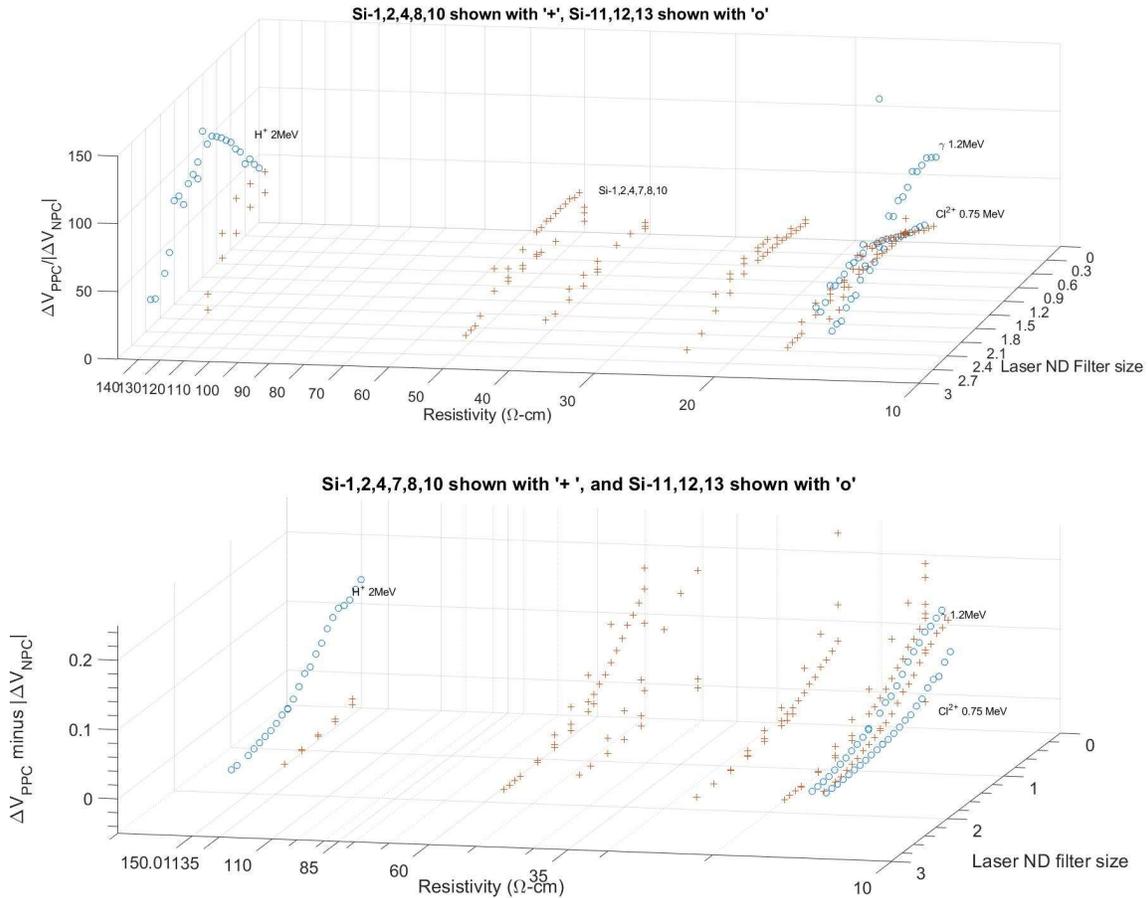

**Figure 17:** (a) shows the ratio of PPC and NPC (magnitude) peak voltages. The $H^+$ and $Cl^{2+}$ sample ratio peaks at ~1.3ND filter size (~0.65 µJ cm$^{-2}$) whereas the gamma-irradiated sample peak ratio gradually increases and becomes maximum ~ 0.3 ND filter size (~5.6 µJ cm$^{-2}$), (b) 3D plot showing the variation of the peak voltage difference to the ND filter size (in reverse order representing fluence) and the resistivity of the normal and irradiated samples.

## 3.3 Variation of PPC and NPC recombination time constants between normal and irradiated samples

### 3.3.1 Single exponential fitting of PPC time constants for normal and irradiated samples

We have taken the PPC transient decay data and fitted it in one exponential $y = ae^{bx}$ for Si-1,11,12, and 13 respectively and plotted '1/b' as a function of laser fluences used during the data acquisition. We note that the proton and chlorine ion beam irradiated samples exhibit a slightly lower time constant in the nano-J to micro-J range. Proton irradiated sample (Si-12) exhibits a low decay period at µJ range whereas, the $Cl^{2+}$ irradiated sample (Si-13) exhibits a longer time constant in the entire range of laser fluences collected here.

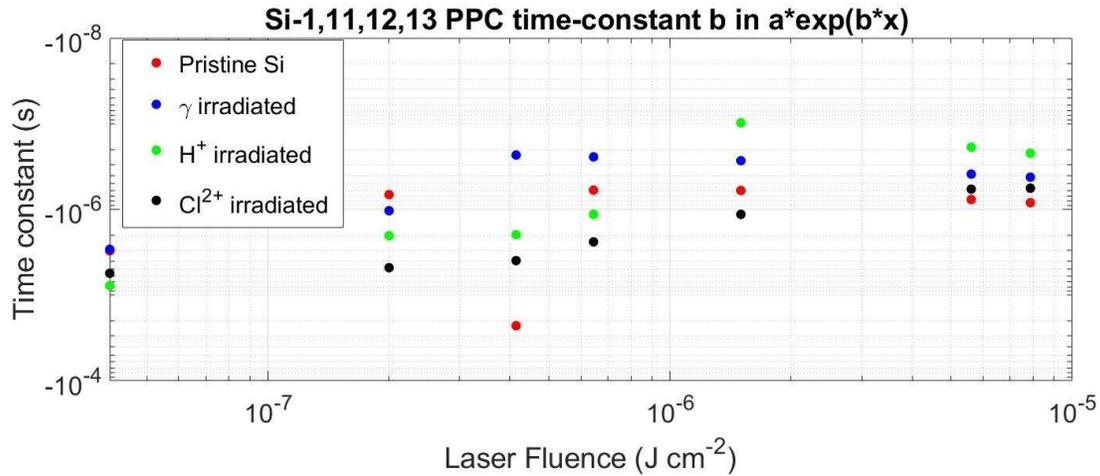

**Figure 18:** Shows the single exponential decay time constant in seconds (1/b) evaluated using the detector RF transients for Si-1, 11, 12, and 13 and shown as a function of the laser fluence.

### 3.3.2 Three exponential fits of PPC and NPC time constants for normal and irradiated samples as a function of fluence

We have fitted the data using the least-squares non-linear fitting available in MATLAB. These are the same transient decay data presented in the previous section 3.2.1 but, with 3 exponential fits representing the function $y = c \exp(-dx) + e \exp(-fx) + g \exp(-hx)$. Ideally, the time constants 'd', 'f', and 'h' represent the radiative, band-to-band ($\tau_1$), trap-assisted Shockley-Read-Hall ($\tau_2$), and the Auger recombination ($\tau_3$) processes in the sample, respectively. By comparing the bar charts appearing for the pristine (Si-1) PPC and NPC (left and right panels of Figs. 19a, 19b, and 19c representing $\tau_1, \tau_2,$ and $\tau_3$ respectively) for the pairs appearing for Si-1, the γ exposed, $H^+$ exposed, and the $Cl^{2+}$ ion beam exposed Si samples. For the case of $\tau_1$, the PPC decay time constants are found to shorten by an order of $10^{-1}$s for the gamma exposure, $H^+$ exposure enhances $\tau_1$ for the highest laser fluence bin (7) and $Cl^{2+}$ exposure also enhances the PPC $\tau_1$ for moderate to high fluence bins. The NPC for $\tau_1$ shows in general a longer time scale and for gamma exposed samples we note a sharp change in time decay time. In general, for $\tau_1$ and $\tau_3$ also, we note that for $H^+$ and $Cl^{2+}$ exposure, both, the PPC and NPC decay time constants are longer (changes from µs to ms range). Notably, the Auger recombination process ($\tau_3$) does not show a negative decay period for NPC decay in the gamma, proton, or chlorine ion irradiated samples unlike for the appearances of negative NPC decay constant ($\tau_1$) for gamma exposed sample and the proton exposed samples (see Fig. 19a right panel numbers 2 and 3) NPC decay time constant $\tau_2$ for the proton exposed samples (see Fig. 19b right panel number 3).

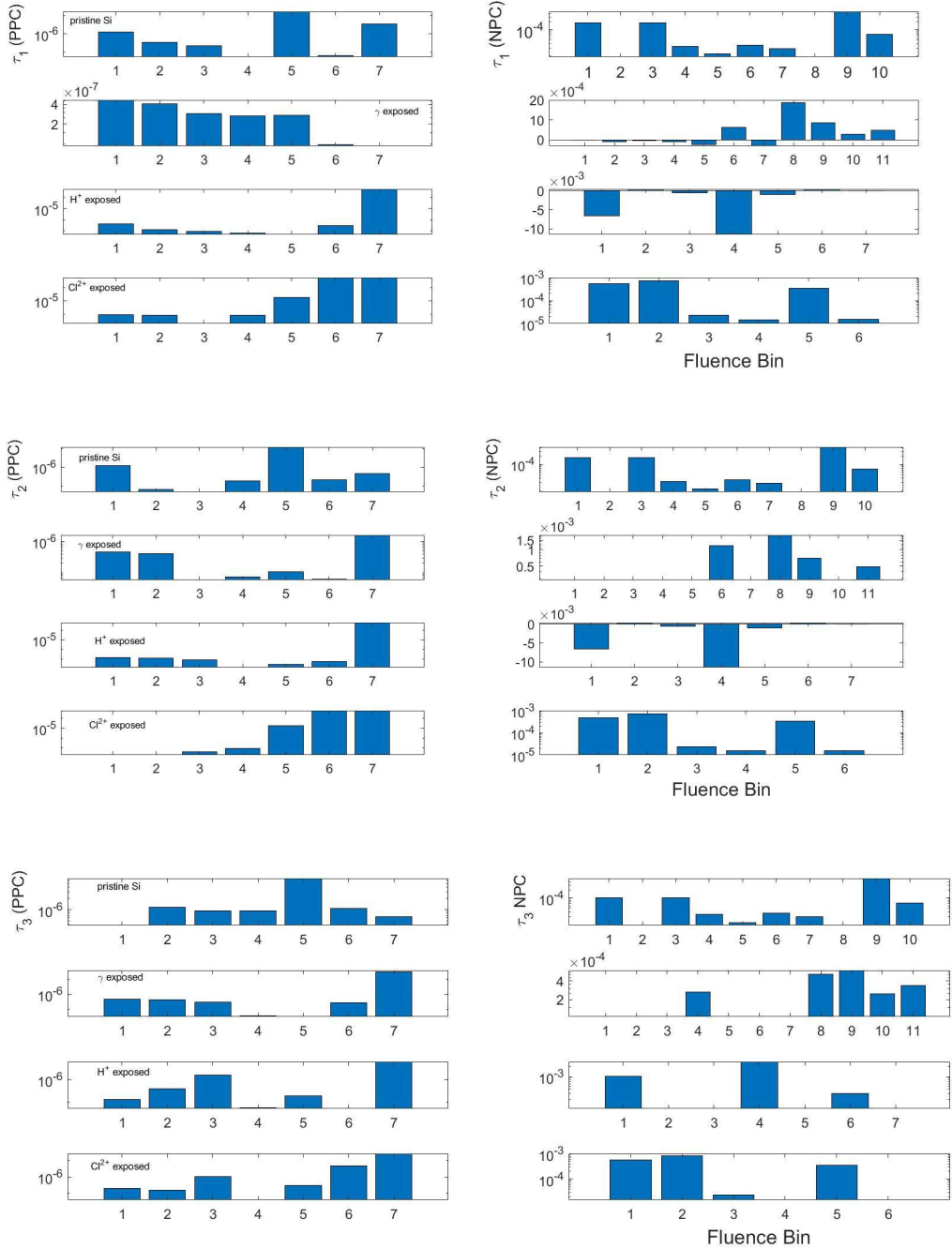

**Figure 19:** (a) The left panel shows the $\tau_1$ from the pristine sample Si-1, the $\gamma$ exposed sample Si-11, the H$^+$ exposed sample Si-12, and the Cl$^{2+}$ exposed sample (Si-13). The right panel showing the same time constant but obtained after fitting the NPC decay signal; (b) is the same as 19(a) except for the trap-assisted time constant $\tau_2$, and (c) is the same as 19(a) and 19(b) except for the Auger process time constant $\tau_3$.

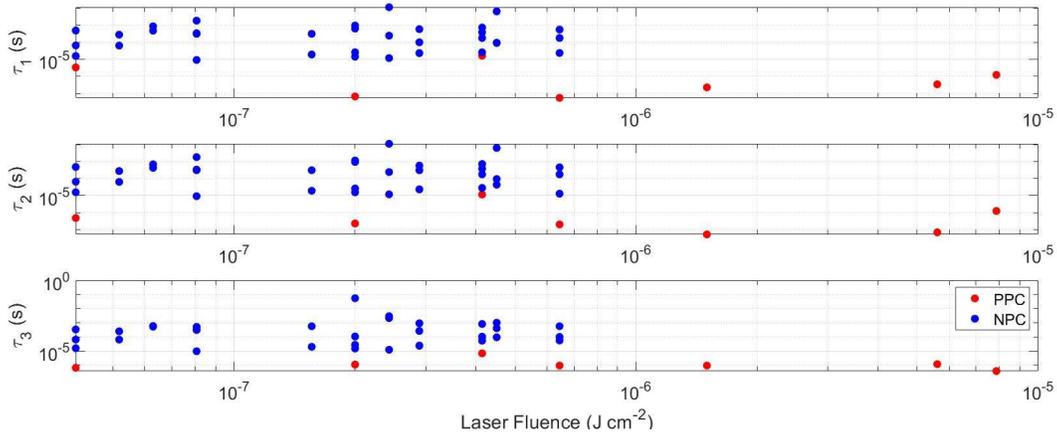

**Figure 20:** Shows the PPC and NPC time constants that are given by laser fluence (shown as fluence bins in Figs. 19a-c) for the Si-1,11,12 and13 combined.

## 4. Summary

We have used 6 pristine (normal samples, 1 N-type, and 5 P-type) moderate resistivity silicon wafers and also compared the normal sample (mostly with Si-1) TRmmWC responses with one gamma-, 1 proton, and one chlorine ion beam irradiated silicon sample drawn from the same parent silicon sample Si-1. Most of the comparisons presented here are obtained for the probe frequency 150 GHz at a very low power level (~ 0.32 mW). The laser excitation is obtained from a narrow pulse 532 nm laser operating at a repetition rate 1kHz. We have studied the TRmmWC response ratio ($\Delta V/V_0$) as a function of laser fluence, flux, the dopant concentration of the sample (resistivity) for the normal and the irradiated samples as well. The parameters compared and results presented pertain to the positive- and negative photoconductivity (PPC and NPC) voltage response (magnitudes) once taking the ratio and another time taking the differences. Following is the summary of the findings in the present analysis of the TRmmWC data for pristine and irradiated Si samples:

- Theoretically calculated transmission coefficient for all samples is highest at 170 GHz in the 110-170 GHz range, and the theoretically computed reflection coefficient is maximum at ~ 130 GHz through the silicon sample (Si-4 used here). We choose the probe frequency 150 GHz due to a modest comparison between the theoretical and experimental reflection coefficients data.
- For the transmitted signal, $Cl^{2+}$ ion irradiated sample exhibits a greater NPC peak voltage that more or less remains constant with laser fluence, $H^+$ irradiated sample traps charge carriers giving rise to an elevated NPC peak that falls off very slowly with laser fluence than the normal (pristine) and the gamma exposed silicon; gamma- and $Cl^{2+}$ irradiated Si does not show change abrupt change in TRmmWC responses as function of sample resistivity unlike the $H^+$ ion beam irradiated sample; PPC peak voltages obtained for 1.3 and 1.6 ND fluence for normal samples maintain peak values almost proportional to laser fluence except for the dopant concentration $8 \times 10^{14}$ cm$^{-3}$; Three different strands are seen

- for PPC and NPC peak voltage ($\Delta V$) and also the TRmmWC response ($\Delta V/V_0$) when they are plotted with the flux (voltage-time product enclosed by the PPC and NPC transients respectively), the knee of the peak and response ratio is observed ~ $10^{-8}$ Wb; irradiated samples exhibit only 2 distinct strands when the response ratio and the peak voltages plotted with flux.
- NPC peak voltage when plotted with PPC peak voltages (either by resistivity or laser fluence) show an exponential dependence on each other; Proton damaged Si exhibits a much higher PPC to NPC peak voltage ratio when the laser fluence is around 1-4 $\mu J\ cm^{-2}$; Irradiated sample peak NPC and PPC departures from the same in Si-1 (parent Si wafer) at a fixed laser fluence (~ 0.41 $\mu J\ cm^{-2}$) show that the 1.2 MeV gamma radiation-damaged sample differences cluster around 0, however, the $H^+$ damaged Si shows a sharp change in the NPC voltage difference than the $Cl^{2+}$ damaged sample. This is probably due to the penetration depth of the $H^+$ beam being much higher than the heavy $Cl^{2+}$ ions.
- Based on the laser fluence-resistivity joint variation plot of the PPC to NPC peak voltage ratio ($\Delta V_{PPC}/\Delta V_{NPC}$) and the difference ($\Delta V_{PPC} - \Delta V_{NPC}$) we note that, in general, the peak voltage ratio for all samples peak around 1.3ND laser fluence corresponding to 0.65 $\mu J\ cm^{-2}$; the gamma-irradiated sample shows a steady rise in the peak voltage ratio and a sharp fall with an increase in laser intensity beyond 0.65 $\mu J\ cm^{-2}$; For the peak voltage difference profiles, it is notable that the $H^+$ irradiated sample shows a sharp change at ~1.5 ND laser ND filter size and, so is seen for the medium resistivity 35-60 $\Omega$-cm normal Si samples.
- 3 exponential fitting: $\tau_1$ PPC decay time constant shortens by 0.1s for gamma exposure when compared with the pristine sample, $H^+$ exposure enhances $\tau_1$ at high laser fluence whereas, $Cl^{2+}$ exposure enhances $\tau_1$ for moderate to very high laser fluences only. For ion exposure both the PPC and NPC decay time constants $\tau_1$ and $\tau_3$ change from μs to ms ranges. Auger recombination ($\tau_3$) does not show a negative decay period for the NPC cases of the gamma- and ion exposed samples; $H^+$ exposure changes the SRH time constants ($\tau_2$) from being positive in the PPC to be negative in the respective NPC voltage responses

One of the main aims of this work was to connect the sample intrinsic parameters such as dopant concentration, thickness, and probe beam penetration depth with TRmmWC measurable charge dynamical parameters, flux, and the laser fluence levels to be able to quantitatively verify the defects leading to TS due to ion beam exposure of silicon samples. We have provided a first-hand preliminary analysis of the 3 experimental datasets and it would be very important to be able to repeat these experiments and with more radiation-damaged samples and by varying the irradiation flux while exposing the samples and then running TRmmWC for a larger frequency band extending from 110 to 325 GHz. A good theoretical treatment of negative photoconductivity in the light of probe-beam-material interaction would be required. It would be also ideal to be able to adopt the 2-defects model for the transient NPC in p-type silicon following the recent idea presented by Zhu et al. (2018) and then be able to compare the experimental data with the modeled outputs.

**Acknowledgments**

The authors acknowledge the support received from UNC General Administration (GA) carbon electronics research program from which the TRmmWC instrument was conceived. BR acknowledges the NSF CREST Grant Award No. HRD-1345219and NSF/PREM grant Award No. DMR-1523617. The authors like to thank Dr. B. Pivac of Ruder Boskovic institute for providing the irradiated Silicon samples, and to Dr. Mark Walters of Duke shared materials instrumentation facility (SMIF) which is supported by NSF Grant No. ECCS-1542015 for allowing 4-probe resistivity measurements. BR also acknowledges partial funding from the Department of Energy National Nuclear Security Administration grant to NCCU No. NA0003979.